Effect of Nano Scale Fe Doping on Superconducting Properties of MgB$_2$

S.X. Dou, S. Soltanian, Y. Zhao, E. Getin, Z. Chen* and J. Horvat
Institute for Superconducting and Electronic Materials,
*School of Materials, Mechanical and Magtronic Engineering,
University of Wollongong, Northfields Ave. NSW 2522 Australia

**Abstract**

Iron is an important sheath material for fabrication of MgB$_2$ wires. However, the effect of Fe doping on the superconducting properties of MgB$_2$ remains controversial. In this work, we present results of nano-scale Fe particle doping in to MgB$_2$. The Fe doping experiments were performed using both bulk and thin film form. It was found that Fe doping did not affect the lattice parameters of MgB$_2$, as evidenced by the lack of change in the XRD peak positions for MgB$_2$. Because of the high reactivity of nano-scale Fe particles, Fe doping is largely in the form of FeB at low doping level while Fe$_2$B was detected at 10wt% doping by both XRD and TEM. There is no evidence for Fe substitution for Mg. The transition temperature decreased modestly with increasing Fe doping levels. The $J_c(H)$ performance was severely depressed at above 3wt% doping level. The detrimental effect of nano-scale Fe doping on both $T_c$ and $J_c(H)$ is attributable to the grain decoupling as a result of magnetic scattering of Fe-containing dopants at grain boundaries.

**Introduction**

The topic of Fe doping into high temperature superconductors (HTS) has been extensively studied and a large number of papers published in the literature. This is because Fe can substitute into the Cu site in cuprates up to 1/3 of Cu without phase segregation. It was also of interest to study the effect of the ferromagnetism of Fe on the superconductivity. There is a fair amount of agreement on the effect on the superconducting characteristics of doping Fe into a given cuprate. A typical example of an HTS compounds for the study of the effect of Fe doping is YBa$_2$Cu$_3$O$_{7-x}$ (YBCO). In this case, the depression of superconductivity is largely attributable to the Cu valence and structure change rather than any magnetic property change as a result of Fe doping [1]. It is interesting to note that the effects on $T_c$ by magnetic (e.g. Fe) and non-magnetic element (e.g. Zn) are comparable for YBCO while the dominating factors are the structure change, oxygen content and Cu site where substitution takes place, on a Cu plane or in a Cu chain.

In contrast, there is little work on the effect of Fe doping on the superconducting behavior of MgB$_2$. This is largely due to the fact that Fe is relatively inert to MgB$_2$ and has been widely used as a sheath material for fabrication of metal-clad MgB$_2$ wire and tapes. However, the magnetism of Fe doping as an additive may have a significant effect on the superconductivity of MgB$_2$. This effect could be more dramatic if the Fe additives are at nano-scale. Jin et al. have studied the effect of Fe addition on $T_c$ and the $J_c - H$ characteristics of MgB$_2$ [2]. They found that 5wt% Fe addition had no effect on $T_c$ and a negative effect on $J_c$. However, it had the least negative effect on $J_c - H$ behavior of all the elements they studied, including Cu, Ag, Y, Ti, and Mo. From their results using Mossbauer spectroscopy and XRD, Kuzmann et al suggested that Fe substitutes into the Mg site in the MgB$_2$ lattice, resulting in modest depression in $T_c$ [3]. These authors did not study the effect of Fe doping on $J_c(H)$. Recently, Prozorov et al. applied a sonication method to produce magnetic Fe$_2$O$_3$ nanoparticles embedded in the MgB$_2$ bulk, resulting in a superconductor-ferromagnetic composite which exhibited enhancement of the magnetic hysteresis due to the improved vortex pinning by the magnetic nanoparticles [4]. In conventional superconductors, submicron magnetic dots in a specially designed pattern have been used as artificial pinning centers in superconductor thin films [5,6]. Because of the important role of Fe as a commonly used sheath material for the fabrication of MgB$_2$ there is an urgent need to clarify the effects of Fe doping on the superconducting properties, in particular on the flux pinning.

In this article, we present results on the effects of nano-scale Fe doping in both bulk and thin film on the lattice structure, $T_c$ and $J_c – H$ characteristics of MgB$_2$. The results from both bulk and thin films demonstrate that nano-Fe doping modestly depressed $T_c$ and severely damaged the $J_c$ in-field performance. There is no evidence to suggest that Fe particles act as pinning centres. We interpret this negative effect of Fe doping on $J_c(H)$ in terms of magnetic scattering by Fe dopants.

**Experimental details**

MgB$_2$ pellet samples were prepared by an *in-situ* reaction method, described in detail previously [7]. Powders of magnesium (99%) and amorphous boron (99%) were well mixed with Fe nanoparticle powder (spherical morphology, size 25nm and surface area 40-60 m$^2$/g) with the atomic ratio of MgB$_2$ and with 0, 1, 5wt% Fe addition. Pellets 10 mm in diameter and 2 mm in thickness made under a uniaxial pressure were sealed in an Fe tube and then heated at 800$^o$C for 30min in flowing high purity Ar, followed by furnace cooling to room temperature.

The Fe doped MgB$_2$ thin films were prepared using a pulsed laser deposition technique with the *ex situ* annealing described previously [8]. In this procedure, a boron precursor film was deposited from a boron target in 10$^{-7}$-10$^{-6}$ Torr vacuum. For Fe doping, a target-switching method was employed. In this method, two targets were mounted onto a carousel. They were a boron target (84% density) and an iron target, respectively. The Fe added B precursor film was achieved by sequentially ablating the B and the Fe target 10 times during the deposition. In each round, the Fe target was ablated for 2 pulses at 1 Hz laser repetition frequency, and the B target was ablated for 30 sec at 10Hz. We calibrated the growth rate of Fe film and B film under our deposition conditions and controlled the number of laser pulses on each target to make the Fe content in the final *ex situ* annealed MgB$_2$ film equal to 3 wt%. The substrate was also kept at 250$^o$C during the deposition. The precursor film was then wrapped with Ta foil and sealed in a stainless steel tube in Ar atmosphere together with Mg pellets. The tube was put into a 900$^o$C furnace and kept for 30min. The heating rate and cooling rates are both about 180$^o$C/min.

Magnetization was measured from 5 to 30 K in magnetic fields up to 9 T using a 9 T Physical Property Measurement System (PPMS, Quantum Design). Bar-shaped samples of about the same size were cut from the as-sintered pellets to minimize size-dependent effects [9]. Magnetic $J_c$ values were determined from the magnetization hysteresis loops using the appropriate critical state model. An empirical magnetic irreversibility line $H_{irr}$ was defined as the field at which $J_c$ falls to 100 A/cm$^2$. Lattice structure and phase compositions were examined using XRD and transmission electron microscopy (TEM) was employed to characterize the morphology of the samples.

**Results and Discussion:**

Fig. 1 shows XRD patterns for the Fe doped and un-doped bulk samples with pure Si as the standard. The XRD pattern for the un-doped sample is consistent with the published indices of MgB$_2$ with a trace amount of MgO. For the Fe doped samples, at an Fe doping level of 2wt% or above, FeB phase is clearly seen as one of main impurities along with MgO and unreacted Mg. The presence of the unreacted Mg is due to the formation of FeB which caused a deficiency of B. Because of the reaction between B and Fe a sample with the nominal composition of Mg$_{0.96}$Fe$_{0.04}$B$_2$ (Fe doping level equal to 5wt%) was made in the same way in order to maintain the stoichiometry of MgB$_2$. However, there is no significant difference between the XRD patterns of this sample and that with 5wt% Fe addition. From the XRD pattern we cannot detect any free Fe presence in the Fe doped samples. To confirm the FeB compound formation during the doping process, we intentionally added more nano-Fe. With a further increase of Fe doping level to 10wt%, in addition to FeB a new compound of $Fe_2B$ was detected.

Fig. 2 shows the transition temperature ($T_c$) for the doped and un-doped samples determined by ac susceptibility measurements. For thin film samples, the $T_c$ showed little difference between the 3wt% Fe doped sample and the un-doped one. For the pellet samples, the $T_c$ obtained as the onset of magnetic screening for the un-doped sample was 38.6K with a transition width of 1.1K. For the doped samples, the $T_c$ decreased with increasing doping level and the transition became broad. $T_c$ drops by 4.6K with a transition width of 7.9K at the Fe doping level of 5wt%. For the sample with the nominal composition of $Mg_{0.96}Fe_{0.04}B_2$ (equal to 5wt% Fe addition) it's the $T_c$ was 35.6 K, 3 K lower than for the un-doped sample. In comparison with substitution cases such as C and Al, the effect of Fe doping on $T_c$ is not too drastic. However, compared with other non-magnetic additives the depression of $T_c$ by Fe doping is slightly stronger. For example, $T_c$ only drops by 1.3K at a doping level of 15wt.% nano SiC in $MgB_2$ [7] and by 1.5K at 10wt% of nano-Si doping [10]. This indicates that the depression of $T_c$ by nano-Fe doping is attributable to magnetic scattering by the ferromagnetic particles.

Fig. 3 shows the $J_c(H)$ curves for the un-doped and Fe-doped $MgB_2$ samples at 20K for different doping levels. It should be noted that the Fe doping not only depressed the zero field $J_c$ but also damaged the $J_c(H)$ behaviour in magnetic fields. At a 5wt% Fe doping level, the $H_{irr}$ (as defined as the field value at $J_c$ = 100A/cm$^2$) dropped from 4.7T to 1.9T at 20K. For the sample with the nominal composition of $Mg_{0.96}Fe_{0.04}B_2$ the $J_c(H)$ performance is better than with 5wt% doping, but follows the same trend. To confirm the negative effect of Fe doping on $J_c(H)$, we present the $J_c(H)$ results of the 3wt% Fe doped $MgB_2$ thin film in Fig. 4. It is clear that the $J_c(H)$ performance in magnetic field was severely depressed due to the Fe doping, consistent with the bulk samples. Because the Fe doped thin films were made using the pulsed laser deposition technique there is no doubt that the Fe particle size is at nanometer scale and Fe distribution is highly uniform within the $MgB_2$ matrix.

**Discussion:**

In contrast to HTS, $MgB_2$ has a relatively large coherence length and small anisotropy. Accordingly the fluxoids to be pinned are string-like and amenable to pinning by particles, precipitates, etc. This opens a window to the success of chemical doping in this material. Many additives (reactive or non-reactive to $MgB_2$), such as Si [10], $Y_2O_3$ [11], Ti, Zr [12], $ZrSi_2$, $WSi_2$, $ZrB_2$, $Mg_2Si$, and $SiO_2$ [13] have been reported to have a positive effect on flux pinning. Under controlled conditions, some substitution elements such as C [14], SiC [7] and Al [15] have shown a significant enhancement in $H_{c2}$ and $J_c(H)$ performance. Only a few dopants, such as Cu, showed any negative effect because of strong reaction with Mg. It is particularly interesting to note that many of these dopants such as SiC, Si, $Y_2O_3$, Ti, and Zr have shown an enhancement effect on flux pinning even though they actively reacted with either B or Mg to form compounds. In comparison, the nano-particle Fe doping has a dramatic depressing effect on $J_c(H)$. Our results are not in agreement with a recent report which claiming that fine $Fe_2O_3$ particles acted as efficient pinning centres to enhance $J_c$ [4]. Our findings also do not support those studies using an array of magnetic dots in low temperature superconductor films [5,6]. It should, however, be pointed out that the latter case is quite different from our experimental conditions. Their Fe dots were designed to be spaced of 600nm apart and to have a dot size of 100nm, while our Fe particles were smaller and the space between them was much smaller and randomly distributed. Also, their thin films were made of Nb and tested in a very low field regime (< 0.15T).

The possible mechanisms for such a negative effect on $J_c(H)$ are proposed to be (1) Fe substitution for Mg which was not found in our study as discussed above; and (2) magnetic scattering of Fe particles that caused depression of superconductivity and weak links at grain boundaries. Kuzmann et al. [3] and Guo et al. [16] suggested possible Fe substitution in the Mg site in the $MgB_2$ lattice [3]. Because these techniques were applied to an entire sample the claim of Fe substitution for Mg was inconclusive. In the present study, we found no convincing evidence of Fe substitution for Mg. If Fe substituted for Mg we would see a change in the lattice parameter as a result of Fe doping because the $Fe^{+3}$ ion radius (0.55A) is much smaller than $Mg^{+2}$ (0.72A). However, we found no change in the lattice parameters with increasing Fe doping level. Fig. 5 shows a comparison between the XRD patterns of the un-doped and 5wt% doped samples. The insets show the (001) and (002) peaks for the two samples. There is no sign of a shift in these peaks even though we used a very slow scan rate. This indicates that there were no changes in *a*- and *c*- axis lattice parameters due to Fe doping. Furthermore, as FeB peaks were clearly visible even at a 2wt% Fe doping level it is evident that the Fe is largely in the form of FeB rather than in the form of any substitution. However, it should be pointed out that we cannot completely rule out Fe substitution for Mg as the current techniques have a limited detection level.

TEM examination and EDS analysis revealed the following features of Fe doped sample. Fe containing particles are uniformly dispersed within the matrix with no indication of agglomerates. Fig 6 shows TEM images of a 5wt% Fe doped sample. In one case round and dark pure Fe particles of 40-50nm size were observed at grain boundaries as shown by the arrow in Fig. 6 (a). Some pure Fe particles of approximately 20nm in size were also observed in this sample (arrows in Fig. 6(b)). FeB particles are also another type of Fe containing particles observed in this experiment as shows for example in Fig. 6 (c). As can be seen, FeB particles are much larger (about 100 nm) and are highly twinned while normal Fe does not have such a high density of twins. The FeB is characterized by electron diffraction and energy dispersion X-ray analysis (EDS). Due to the large difference between the size of pure Fe particles and the size of FeB particles and the small number of pure Fe particles, the volume fraction of FeB would be significantly higher than pure Fe, which is consistent with the results of XRD where only FeB is detected. EDS analysis of the $MgB_2$ grains showed a small peak of Fe, but this again may be from Fe containing compounds as the grain size of $MgB_2$ is very small, about 100nm. Thus far, the evidence for Fe substitution for Mg remains inconclusive.

Because of the large coherence length of $MgB_2$ many nano-particles at grain boundaries do not cause weak links but act as effective pinning centres which has been well demonstrated by a number of doping studies [7, 9-15]. In the case of Fe doping, FeB particles at grain boundaries cause grain decoupling because FeB particle scattering affects a much larger volume surrounding the particle than nonmagnetic particles. Thus, the superconductor volume decreased with increasing Fe doping level, resulting in the reduction in zero field $J_c$. With larger amounts of Fe, the layer decoupling the grains was thicker, Fe containing compounds become weak links rather than flux pinning centres. This is a typical characteristic for HTS in which weak links are the dominating factor for current flow. We do not have this weak link characteristics even at higher doping levels for non-magnetic dopants [16]. 20%SiC addition to $MgB_2$ not only showed no such weak link feature but had a strong enhancement effect on $J_c(H)$ [17]. Especially we should stress that SiC also strongly reacted with Mg to form $Mg_2Si$ and all the additives and reaction products stay at grain boundaries. It is evident that $MgB_2$ is highly tolerant of nonmagnetic impurities but susceptible to ferromagnetic impurities.

Although the Fe doping showed a negative effect on both $T_c$ and $J_c(H)$ this negative effect may not be too drastic in an Fe sheath situation as the contact area between the Fe sheath and the $MgB_2$ is much smaller than the case of nano-Fe particle doping. The reaction between Fe and B is only limited in the interface region. Moreover, the reactivity of Fe is much lower than that of many elements such as Cu, Ag, Ti, etc. In addition, the ferromagnetic property of the Fe sheath provides a magnetic shielding effect that enhances the $J_c(H)$ performance in a low field regime as reported previously [18]. This positive effect of magnetic shielding counterbalances the small negative effect of the Fe sheath on $J_c(H)$.

## Conclusion

In summary, we have demonstrated that nano-scale Fe particle doping depressed both $T_c$ and $J_c(H)$ in both bulk and thin film samples. Because of their high reactivity, in the in-situ process the nano-scale Fe particles reacted with B to form FeB and $Fe_2B$ which were homogenously distributed within the matrix of bulk and thin film $MgB_2$. Fe substitution for Mg is unlikely but remains inconclusive. The strong depression in $J_c(H)$ performance by nano-Fe particle doping is attributable to the decoupling effect of Fe-containing particles within the grains and at grain boundaries.

## Acknowledgement

The authors thank M. Ionescu, M.J. Qin, A.V. Pan and T. Silver for their helpful discussion on this work. The work was supported by the Australian Research Council, Hyper Tech Research Inc, OH, USA, and Alphatech International Ltd, NZ.

**Figure Captions:**

Figure 1. XRD patterns for the Fe doped and un-doped samples. Pure Si is used as standard.

Figure 2. Transition temperature ($T_c$) for the doped and undoped samples determined by ac susceptibility measurements: (a) thin film and (b) pellet.

Figure 3. $J_c(H)$ curves for the undoped and Fe-doped $MgB_2$ samples at 20K for different doping levels.

Figure 4. $J_c(H)$ curves for the undoped and 3wt% Fe doped $MgB_2$ thin film.

Figure 5. Comparison between XRD patterns of the un-doped and 5wt% doped samples. The insets show the (001) and (002) peaks for the two samples.

Figure 6. TEM images of 5wt% Fe doped sample, showing round and dark particles of 40-50nm size present at grain boundaries (a). Some particles of 20nm size are present within grains (b). Fig. 6 (c) shows a FeB single crystal, containing a high density of twins.

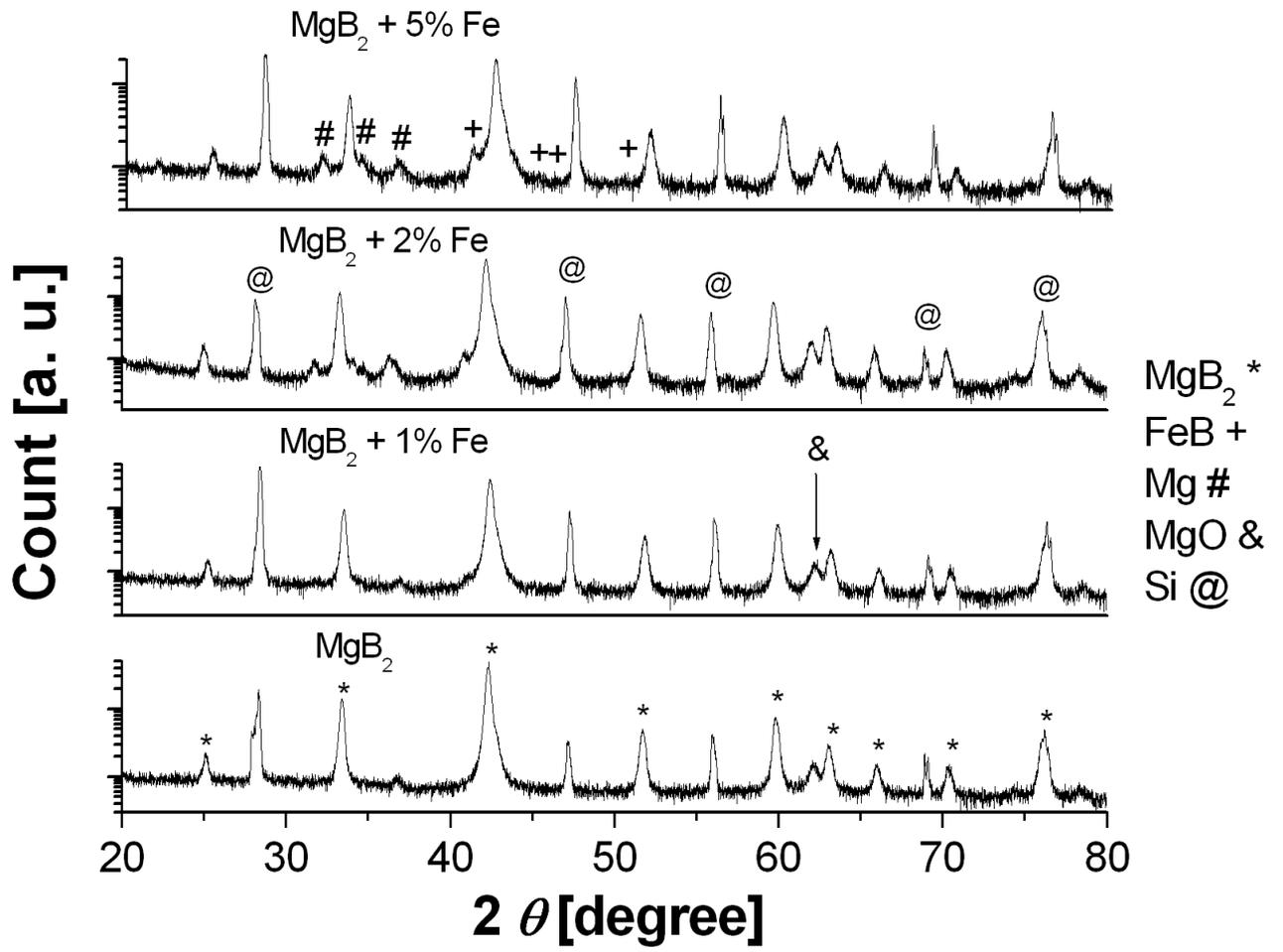

Fig. 1

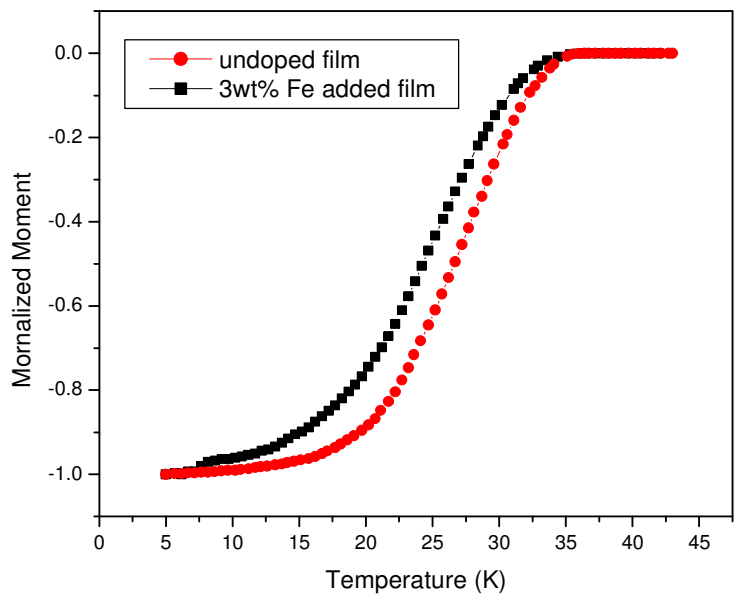

Fig. 2(a)

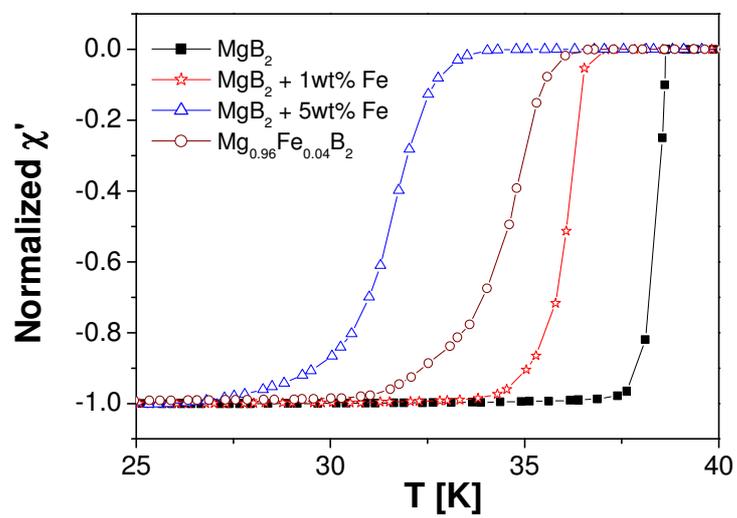

Fig. 2(b)

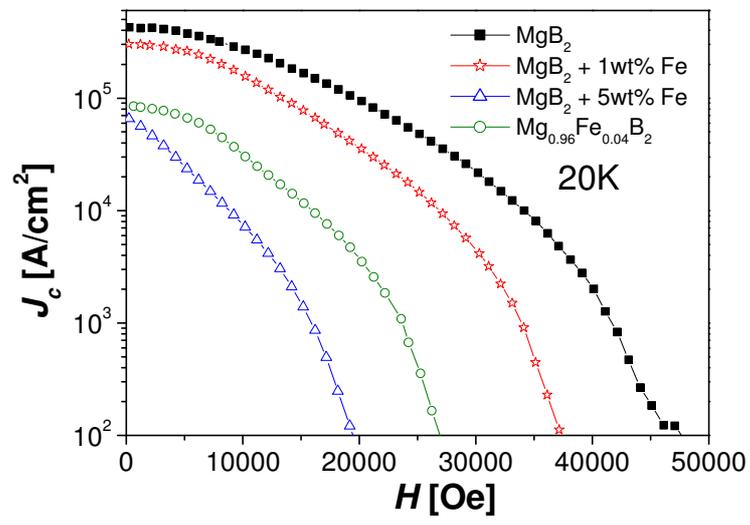

Fig.3

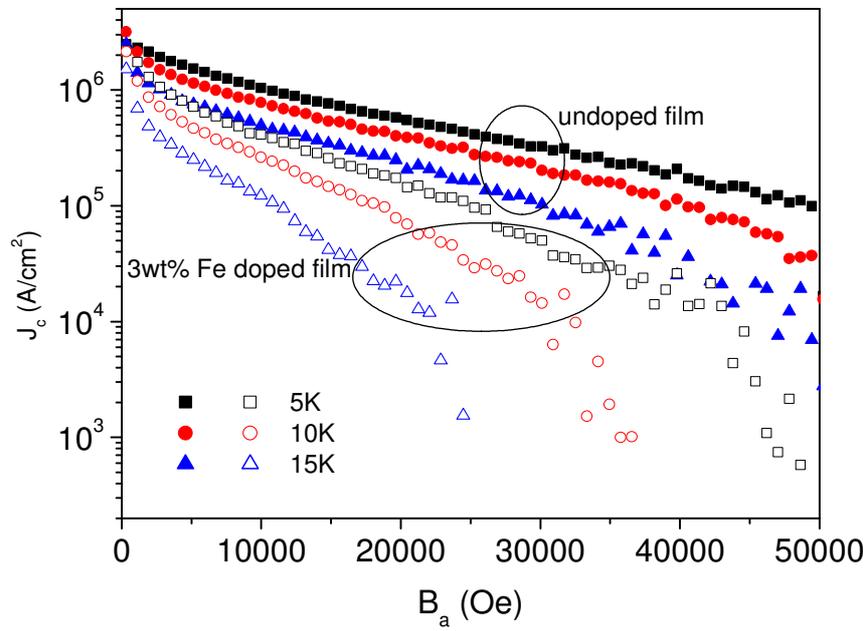

Fig. 4

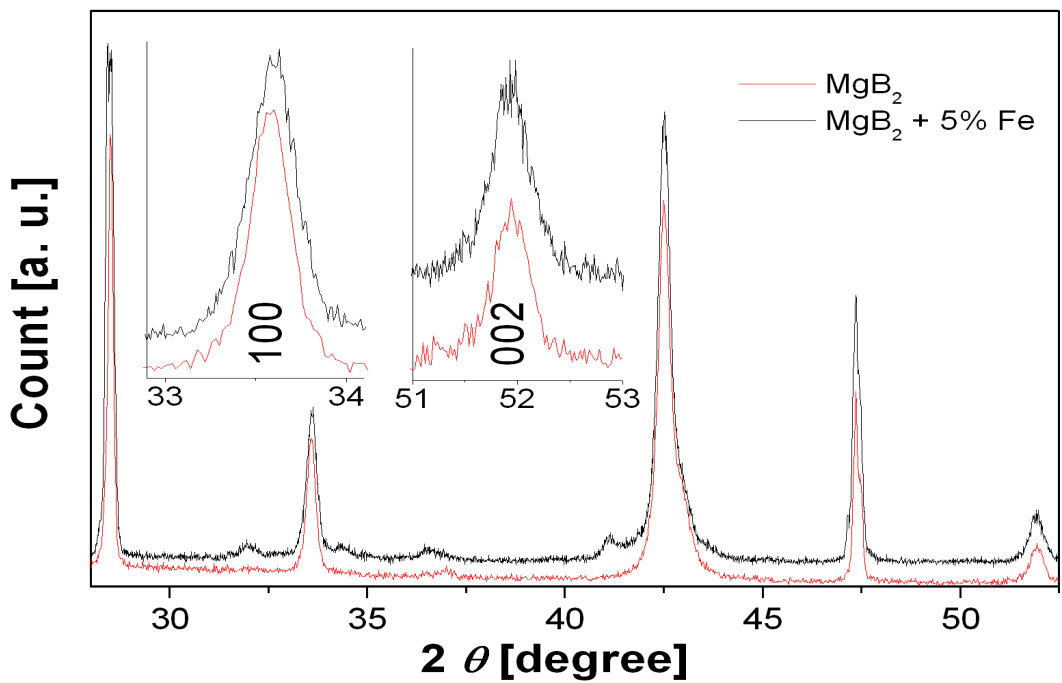

Fig. 5

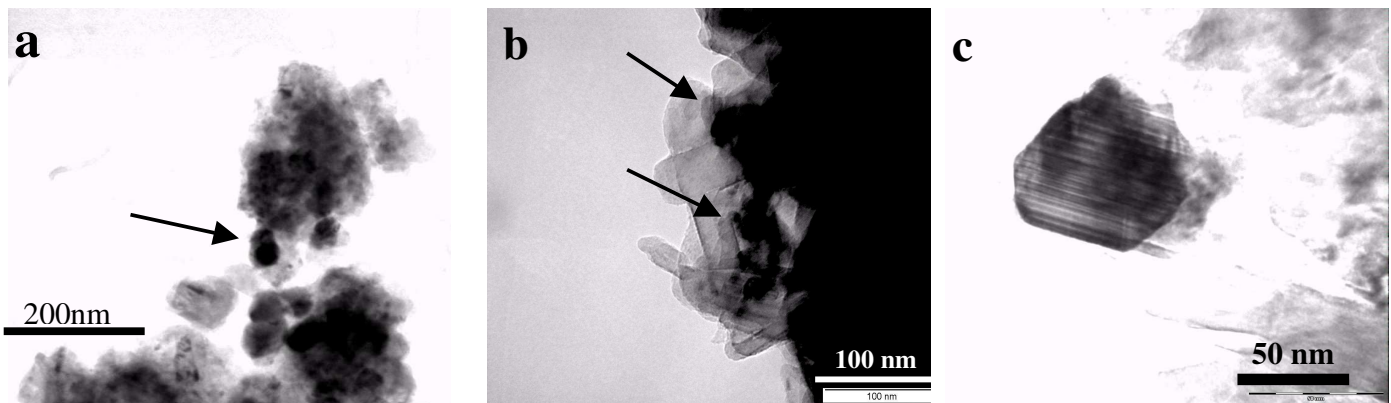
Fig. 6